\documentclass[aps,twocolumn,showpacs,floatfix]{revtex4}
\usepackage{graphicx}
\usepackage{epsf}
\usepackage{subfigure}
\usepackage{epstopdf}
\usepackage{color}

\begin{document}

\title{Quantum thermal equilibration from  equipartition}

\author{A. V. Ponomarev$^{1}$, S. Denisov$^{1}$, P. H\"{a}nggi$^{1}$, and J. Gemmer$^{2}$}
\address{$^{1}$ Institute of Physics, University of Augsburg, Universit\"atstr. 1, D-86159 Augsburg, Germany}
\address{$^{2}$ Fachbereich Physik, Universit\"at Osnabr\"uck - Barbarastrasse 7,
D-49069 Osnabr\"uck, Germany}

\date{\today}

\begin{abstract}
The problem of mutual equilibration between two finite, identical quantum systems,
$A$ and $B$, prepared initially at {\it different} temperatures is elucidated.
We show that the process of energy exchange between the two systems  leads to
accurate equipartition within energy shells in the Hilbert space of the total
non-interacting, composite system, $A \otimes B$. This scenario occurs under the
general condition of a {\it weak interaction} between the systems. We predict that the
sole hypothesis of such equipartition  is sufficient to obtain a relaxation of the peers, $A$ and $B$,
towards a common thermal-like state. This conjecture is fully corroborated by an exact diagonalization of
several quantum models.
\end{abstract}

\pacs{05.30.-d,03.65.Aa}

\maketitle

\section{Introduction} The time evolution of an isolated quantum system after applying a sudden change for one of its parameters, i.e., -- a quench -- has recently gained considerably attention, both in the theoretical and experimental physics communities~\cite{polkovnikov}. State of art numerical simulations~\cite{kollath,manmana,rigol,eckstein,igloi,fine,ates}, motivated by recent advances in manipulations with ultracold atoms~\cite{bloch}, have not only allowed to validate a number of theoretical predictions~\cite{sred,deutsch,gemmer}, but also produced several conceptually new research directions. One of these tracks refers to the exploration of the quench machinery as an effective tool to drag the system of interest into a new state. The latter can effectively mimic the state of thermal equilibrium -- without the need of coupling the system to a heat bath~\cite{rigol1}. `Mimic' means here that the expectation values of relevant observables are close to those following from the thermal Gibbs state, $\varrho_T \propto \exp(-\beta H)$, $\beta = 1/k_\mathcal{B}T$.

The equilibration between two identical, initially non-interacting systems, $A$ and $B$, can be considered as a quench applied to the composite system,
\begin{equation}
\label{eq:setup}
H^\lambda = H_{A} \otimes \mathbf{1}_{B} + \mathbf{1}_{A}\otimes H_{B} + \lambda(t) H^{\rm int},
\end{equation}
starting out from the noninteracting limit, $\lambda = 0$, to the regime of interaction, $\lambda = \lambda_{\rm int}$.
It has been shown with prior work~\cite{ponomarev11} that for the initial product state, prepared at different temperatures, $T_A$ and $T_B$, $\varrho(0)=\varrho_{T_{\rm A}}^{\rm A} \otimes \varrho_{T_{\rm B}}^{\rm B}$, the step-like quench $\lambda(t) = \lambda_{\rm int}\theta(t)$ evolved the composite system into a new state, $\varrho(t)$, such that, for the times $t > t_{\rm eq}$, the reduced density matrices, $\varrho^{\rm A}(t)$ and $\varrho^{\rm B}(t)$, become quasistationary~\cite{quasi} and mimic perfectly a thermal equilibrium with a common temperature $T_{\rm eq}$. Although this scenario seemingly is universal, in a sense that it works equally well for very different physical systems, the physical mechanism at work remained elusive.

With this study we address this open problem. We show that mutual thermal relaxation of two
finite quantum systems follows from a generic hypothesis about the asymptotic state of the
composite system after application of a weak interaction quench: namely, {\it  equipartition inside  energy shells}
$E = \epsilon^{A} + \epsilon^{B}$ of the identical spectra of the composite system $H_{A} \otimes H_{B}$
constitutes a sufficient condition for the emergence of the \textit{mutual thermal equilibration} between the
system's halves. We corroborate this conjecture by using four different types of models, including synthesized
Hamiltonians with different distributions of energy levels and a system of two interacting spin clusters.

\section{Setup} The model (\ref{eq:setup}) consists two identical quantum systems, $A$ and $B$, with identical
finite spectra, $\{\epsilon_j\}$, $j=0,...,N-1$, of width $\Delta\epsilon=\epsilon_{N-1}-\epsilon_0$,
and a set of eigenstates $\{\vert j\rangle \}$. The corresponding energy level distribution is encoded by the density of states,
\begin{equation}
\label{eq:density}
n(\epsilon) = \sum_{j=0}^{N-1}\delta(\epsilon-\epsilon_j).
\end{equation}
The initial states of the systems are given by Gibbs density matrices, $\varrho^{\rm A}(0) = \varrho_{T_A}$
and $\varrho^{\rm B}(0) = \varrho_{T_B}$, at the temperatures $T_A$ and $T_B$.
The initial state of the total composite system in the product basis, 
$\{ \vert \Psi^0_{m(i,j)} \rangle = \vert j\rangle \otimes \vert k\rangle\}$,
$E_{m(j,k)} = \epsilon_j + \epsilon_k$,
is represented by a diagonal density matrix,
$\rho^{\rm tot}(0) = \rho^{\rm tot}_{jk,j'k'}=\rho^A_{j,j'}\rho^B_{k,k'}=\delta_{j,j'}\delta_{k,k'}p_{jk}^0$, $p^0_{jk} = \exp(-\epsilon_j\beta_A-\epsilon_k\beta_B)\mathcal{Z}_A^{-1}\mathcal{Z}_B^{-1}$, where $\mathcal{Z}_{A,B}$ are the partition functions, $\mathcal{Z}_{A,B}=\int_{-\infty}^{\infty} d\epsilon \exp(-\epsilon\beta_{A,B})n(\epsilon)$. Henceforth, we set $\epsilon_0=0$ and use ${\Delta\epsilon}$ as the energy unit if not specified otherwise.

\section{Equilibration induced by equipartition}
We define the energy shells of the composite system in the product
basis $\{\vert \Psi^0_m \rangle\}$ by using the condition $|E_m - E| < \delta$~\cite{bol}.
The constant $\delta$ is chosen small with respect to the spectral width, $\delta \ll {\Delta\epsilon}$,
but still larger than the mean level spacing of the composite system,
$\delta \gg 2\Delta\epsilon/(\mathcal{N}-1)$, with $\mathcal{N}=N \times N$ energy levels.
The last condition implies that the every energy shell contains many eigenstates. The switch-on of
an interaction Hamiltonian, $\lambda_{\rm int} H^{\rm int}$, which is non-diagonal in the product basis,
generates a set of new eigenstates,
$\vert \Psi_m \rangle$: $H^\lambda\vert \Psi_m \rangle=E_m^{\lambda}\vert \Psi_m \rangle$.
If we sort both sets of eigenstates, $\{\vert \Psi^0_m\rangle\}$ and $\{\vert \Psi_m\rangle\}$,
with respect to their energies, $E_m$ and $E_m^{\lambda}$, we obtain a bell-shaped overlap
function $f_{m'}(m)=\vert\langle \Psi^0_{m'} \vert \Psi_{m'+m}\rangle\vert^2$,
centered at $m'$~\cite{deutsch}, with a width that grows with the strength of
perturbation $\lambda_{\rm int}$~\cite{kolovsky06, kota06}. Throughout this study we 
assume the weak coupling limit, 
\begin{equation}
\label{eq:weak coupling}
\lambda_{\rm int} (\epsilon^{\rm int}_{\mathcal{N}-1}-\epsilon_0^{\rm int}) \ll \Delta \epsilon \;,
\end{equation}
obeying, in addition, the condition 
\begin{equation}
\label{eq:not-to-weak coupling}
\lambda_{\rm int} \| H^{\rm int} \rho^{\rm tot}(0) \| > \bar{s}_{\rm tot}=2\Delta \epsilon/(\mathcal{N}-1),
\end{equation}
where $\| ... \|$ is the operator norm in the Hilbert space of the composite system and $\bar{s}_{\rm tot}$ is the mean level-spacing. 
The last condition means that the interaction should not be \textit{too weak}, otherwise the non-thermal scenario of arithmetic-mean
equilibration \cite{ponomarev11} would take place.

In common setups of quench studies the isolated system is initially prepared in a $m$-th eigenstate
(typically in its ground state, $\vert \Psi^0_0\rangle$~\cite{kollath, manmana, rigol, eckstein, rigol1})
of the Hamiltonian $H^{\lambda=0}$. A weak quench then results in a local smearing of the initial wave
function over the narrow set of new eigenstates, given by the function $f_{m'}(m)$, so that `microcanonical thermalization'
can be expected~\cite{rigol, deutsch, sred, jensen, tasaki}.
Microcanonical thermalization implies that a closed quantum system is transformed into a new state, which satisfy
Boltzmann's postulate of equal {\it a priori} probability~\cite{bol}, here applied to the quantum states within an
energy shell~\cite{esposito}. Evidence that this may indeed be expected under quite general conditions~\cite{vonneumann, bocchieri, ates}
is nowadays discussed under the label ``quantum typicality''~\cite{goldstein, reimann}.

We start by extending the concept of  thermalization to the case of the bipartite system initially prepared in the product state $\varrho(0)$. At time $t = 0$ we turn on the quench by setting $\lambda \neq 0$. Then, after some elapsed characteristic time $t_{\rm rel}$, we switch-off the perturbation and investigate the state of the system with respect to the product basis $\vert \Psi^0_m \rangle$. By representing the system Hilbert space, sheared by energy shells of different energies $E$, as having an onion-like structure, we conjecture that a proper perturbation will initiate the population exchange between the eigenstates within each shell, -- independently of the remaining part of the system Hilbert space~\cite{gemmer1}. This exchange will lead finally to the equipartition of the level populations within each energy shell.

\begin{figure}[t]
\includegraphics[width=0.45\textwidth]{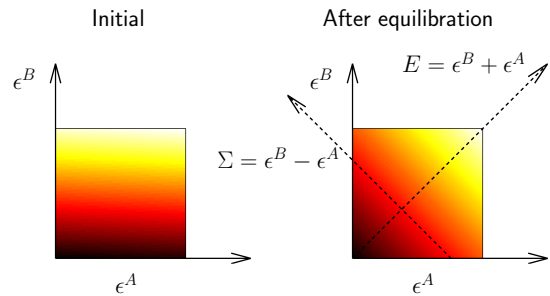}
\caption{(color online). Thermal equilibration between two finite, identical quantum systems following the  equipartition scenario. Systems, initially prepared at different temperatures (left), eventually arrive after relaxation at a  quasi-equilibrium state (right), characterized by distributions uniform along the $\Sigma$-axis (equipartition). The color coding (decreasing in weight from bright to dark) depicts the behavior of the pdf $P[\epsilon^A, \epsilon^B](t)$, see Eq.~(\ref{eq:pdf}). Note that the equipotential lines on the left panel are not horizontal but slightly inclined. The inclination is small due to a large difference between the peer's initial temperatures, $T_A \gg T_B$.} \label{fig1}
\end{figure}

In order to cast our hypothesis into a formal mathematical language we first introduce a
two-dimensional probability density function (pdf):
\begin{equation}
\label{eq:pdf} P[\epsilon^A, \epsilon^B](t) = \sum_{j,k = 0}^{N-1} p_{kj}(t) \delta(\epsilon^A -\epsilon_j)\delta(\epsilon^B -\epsilon_k),
\end{equation}
where the populations $p_{jk}(t)$ are governed by the diagonal elements of the
total density matrix $\rho_{jk,j'k'}^{\rm tot}(t)$.
The initial pdf is given
by $P[\epsilon_i^A, \epsilon_j^B](0)= \exp(-\beta_A \epsilon_i^A - \beta_B \epsilon_j^B )\mathcal{Z}_A^{-1} \mathcal{Z}_B^{-1}$,
see Fig.~\ref{fig1}.
It is useful to introduce the auxiliary variables,
$E=\epsilon^A+\epsilon^B$ and $\Sigma=\epsilon^A-\epsilon^B$, which form a
new coordinate axes. The first variable, $E$, defines the above mentioned energy shell,
while the second one, $\Sigma$, can be used to label the states within the shell.
In this representation the initial condition assumes the form
$p^0_{jk} = P^0(E_m,\Sigma_m)=\exp(-E_m\beta^{+}-\Sigma_m\beta^{-} )\mathcal{Z}_A^{-1} \mathcal{Z}_B^{-1}$,
with the two inverse temperatures $\beta^{+} = (\beta_A+\beta_B)/2$ and $\beta^{-} = (\beta_B-\beta_A)/2$.
The density of states in new variables, $\bar{n}(E,\Sigma)$,  does generally not factorize.

According to the proposed equipartition scenario, after equilibration the diagonal
elements of the total system density matrix, $P^{\rm eq}(E,\Sigma)\equiv P^{\rm eq}(E)$,
derive from the equipartition of the probability over the corresponding energy shells, reading
\begin{equation}
\label{eq:step2}
P^{\rm eq}(E) = \frac{e^{-\beta^{+}E}}{\mathcal{Z}_A\mathcal{Z}_B}\frac{\int_{-\eta(E)}^{\eta(E)} \bar{n}(E,\Sigma) e^{-\beta^{-} \Sigma }d\Sigma}{\int_{-\eta(E)}^{\eta(E)}\bar{n}(E,\Sigma)d\Sigma},
\end{equation}
where the integration limits are $\eta(E) = E$ for $0\leq E \leq \Delta\epsilon$,
and $\eta(E) = 2\Delta\epsilon-E$ for $\Delta\epsilon\leq E \leq 2\Delta\epsilon$, see Fig.~1.
Note that the expression (\ref{eq:step2}) conserves energy within a specific shell. Then the energy level populations of a single peer can be evaluated as:
\begin{equation}
\label{eq:step3}
p^{\rm eq}_j = \int_0^{\Delta\epsilon} P^{\rm eq}(\epsilon_j+\epsilon) n(\epsilon)d\epsilon.
\end{equation}
Note that the equilibrium distribution explicitly depends on the density of states, $n(\epsilon)$, of the system Hamiltonian. Below, by using three different classes of system Hamiltonians we demonstrate that (i) the smearing along the $\Sigma$-axis is sufficient in producing thermal equilibration between the peers and in fact (ii) such smearing indeed is achieved in those systems after an interaction quench.

\begin{figure}[t]
\center
\includegraphics[width=0.45\textwidth]{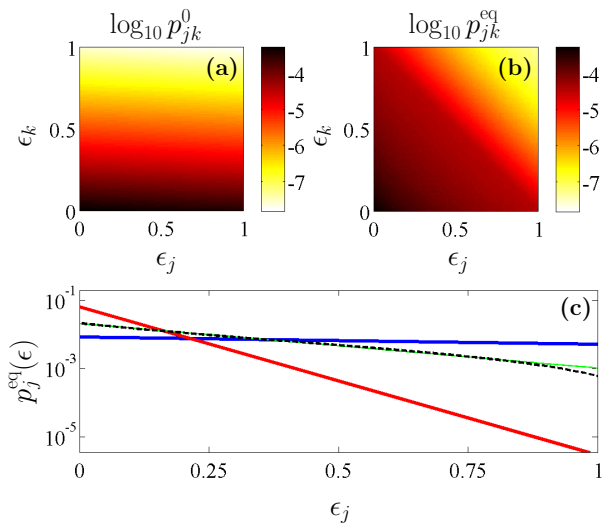}
\caption{(color online). Equilibration between two identical finite quantum systems with $N = 151$ uniformly distributed energy levels. In panel (a) we depict the diagonal elements of the total system density matrix before the interaction quench and in panel (b) the result after equilibration occurred. Notice full agreement with the equipartition scenario, see Fig.~1. The initial temperatures are $k_\mathcal{B}T_A=\beta^{-1}_A=2\Delta\epsilon$ and $k_\mathcal{B}T_B=\beta^{-1}_B=0.1\Delta\epsilon$. Panel (c)~ depicts the populations before and after equilibration. The initial populations, $p^A_i(0)$ and $p^B_i(0)$, are denoted by the thick solid (blue for system $A$ and red for system $B$) lines. The resulting equilibrium populations (thick dashed line) agree (within line thickness) with the analytical prediction, Eq.~(\ref{eq:step3}), and are very close to the canonical thermal populations obtained from the energy conservation condition, Eq.~(\ref{eq:temp_eq}) (thin (green) line). Energy is measured in units of the spectral width~$\Delta\epsilon$.} \label{fig2}
\end{figure}

In order to validate our predictions we performed calculations for three different classes of synthesized Hamiltonians, with uniform, semicircular and a Gaussian density of states, Eq.~(\ref{eq:density}). Finally, we investigated the thermal equilibration between two finite spin clusters.

\section{Thermal equilibration between peers with uniform distributions of energy levels} We have synthesized a 
Hamiltonian with $N=151$ levels, distributed them randomly and uniformly in the interval $[0,\Delta\epsilon]$. 
The interaction Hamiltonian is composed as the product of two identical matrices, $H^{\rm int} = Y_{A} \otimes Y_{B}$, 
where the matrix $Y = Y_A = Y_B$ has been drawn from a Gaussian Orthogonal Ensemble (GOE). Namely, $Y = (R+R^{\rm T})/2$, 
where the matrix $R$ in the product basis is given by its real elements obeying standard normal distribution \cite{edelman}. 
The interaction between the peers is within the weak-coupling limit, so that the interaction quench does not cause appreciable 
heating of the composite system, but still is strong enough as to guarantee the thermal-like equilibration scenario~\cite{ponomarev11}. 
Here we use the dimensionless coupling constant $\lambda_{\rm int} = 0.015 (\overline{s}/\overline{h})$, 
where $\overline{s}=\Delta\epsilon/(N-1)$ is the mean level spacing, and $\overline{h}=\sum_{m,m'=1}^\mathcal{N}\vert H_{m,m'}^{\rm int}\vert/\mathcal{N}^2$.

The initial and the equilibrium population pdf's, as obtained by the exact diagonalization of the composite system 
with $\mathcal{N} = N \times N = 22801$ states, are presented with Fig.~\ref{fig2}. 
The equilibrium pdf $P(E,\Sigma)$ shown in Fig.~\ref{fig2}(b) assumes a stripe-like structure, 
being uniform along the $\Sigma$-axis, in full agreement with the equipartition scenario, see Fig.~\ref{fig1}. 
We also checked that the emerging equilibrium populations $P^{\rm eq}(E)$ follow closely the predicted result in Eq.~(\ref{eq:step2}).

The equilibrium values of $p_j^{\rm eq}$ for a single peer obtained by using Eq.~(\ref{eq:step3}) are shown in Fig.~\ref{fig2}(c) 
by the dashed thick line. The analytical prediction in Eq.~(\ref{eq:step3}) are indistinguishable from the numerical data 
points obtained from the direct diagonalization of the composite Hamiltonian in Eq.~(\ref{eq:setup}). 
Except for some small deviation in the high-energy tail, both distributions fit almost perfectly the thermal 
distribution with the equilibrium temperature $T_{\rm eq}$ extracted from the condition of energy conservation~\cite{ponomarev11},
\begin{eqnarray}
\label{eq:temp_eq}
\sum_k\epsilon_k\frac{e^{-\epsilon_k/k_\mathcal{B}T_{\rm eq}}}{Z_{\rm eq}}
= \sum_k\left[\epsilon_k\frac{e^{-\epsilon_k/k_\mathcal{B}T_A}}{2 Z_A}
+ \epsilon_k\frac{e^{-\epsilon_k/k_\mathcal{B}T_B}}{2 Z_B}\right],
\end{eqnarray}
see the thin solid (green) line in Fig.~\ref{fig2}(c).

\section{Semicircular distribution of energy levels} In the present example we use spectra that are typical for the class of Hamiltonians modeled by a random matrix drawn from GOE~\cite{haake}. In the limit of large number of levels, the density of states of a single peer can be approximated by the continuous semicircular distribution, $n(\epsilon)=(4/\pi)\sqrt{1/4-(\epsilon-1)^2/4}$, see the inset in Fig.~\ref{fig3}(a). Thus, for the total system we have $\bar{n}(E,\Sigma)=1/\pi\sqrt{(E^2-\Sigma^2)^2-4(E^2-\Sigma^2)(E-1)}$. The results of the exact diagonalization perfectly match the prediction in Eq.~(\ref{eq:step3}). As for the first example, the thin solid (green) line indicates the distribution with the equilibrium temperature given by Eq.~(\ref{eq:temp_eq}).

\section{Model with Gaussian distribution of energy levels} This next class of Hamiltonians refers to quantum systems possessing a finite number of interacting particles or spins, as realized with fermionic~\cite{thesis} and bosonic Hubbard models~\cite{kollath11}. In the limit $N \longrightarrow \infty$ the corresponding density of states can be approximated by the continuous Gaussian function $n(\epsilon)\propto \exp\left[-(\epsilon-1/2)^2/(2\sigma^2)\right]$, wherein both the width $\sigma$ and energies $\epsilon$ are in units of the total width $\Delta\epsilon$.

\begin{figure}[t]
\includegraphics[width=0.45\textwidth]{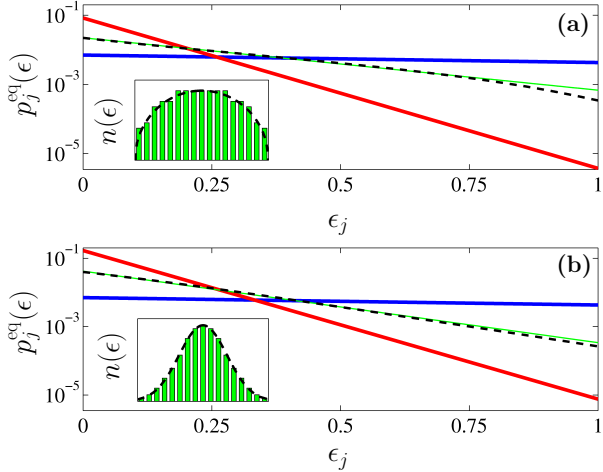}
\caption{(color online). The same as in Fig.~\ref{fig2} but here for the equilibration scenario between two identical peers with a semicircle (a) and Gaussian (b) density of states.
The variance of the Gaussian distribution is $\sigma = 0.1833 \Delta\epsilon$. The remaining parameters are the same as in Fig.~\ref{fig2}(c). Insets: exact semicircle and Gaussian distributions, dashed (black) lines, and the density of states of the finite synthesized Hamiltonian with $N=181$ energy levels (histograms).} \label{fig3}
\end{figure}

In distinct contrast to the semicircle distribution, the Gaussian density of states remains factorized after the frame transformation, $\bar{n}(E,\Sigma)\propto\exp\left[-(E-1/2)^2/(2\sigma^2)\right]\exp\left[-\Sigma^2/(2\sigma^2)\right]$. Therefore, Eq.~(\ref{eq:step2}) reduces (up to irrelevant normalization constant) to the form:
\begin{equation}
\label{eq:step2c}
P^{\rm eq}(E) \propto e^{-\beta^{+}E}\frac{\int_{-\eta(E)}^{\eta(E)}e^{-(\Sigma^2+\sigma^2\beta^{-}\Sigma/2) /(2\sigma^2)}d\Sigma}{\int_{-\eta(E)}^{\eta(E)}e^{-\Sigma^2/(2\sigma^2)}d\Sigma}.
\end{equation}
In the limit of a very broad Gaussian distribution, $\sigma\gg\Delta\epsilon$, the above expression approaches the foregoing result of a uniform distribution, see Fig.~\ref{fig2}. In the opposite limit of a very narrow distribution; i.e., $\sigma \ll \sqrt{\Delta\epsilon/\beta^{-}}$, the integrals in the numerator and denominator of Eq.~(\ref{eq:step2c}) yield approximately the same values, thus rendering the Boltzmann-like distribution,
\begin{equation}
P^{\rm eq}(E) \propto e^{-\beta^{+}E},
\end{equation}
for the composite system.
This limit corresponds to a ``strong thermalization''
numerically observed with two coupled Bose-Hubbard models~\cite{zhang11}.
Accordingly, both peers also relax to the thermal states of the same temperature,
$T_{\rm eq} = (k_\mathcal{B}\beta^{+})^{-1}$, see Fig.~\ref{fig3}(b).
It is noteworthy that the strong thermalization was absent in the
previously considered cases.

\begin{figure}[t]
\includegraphics[width=0.34\textwidth]{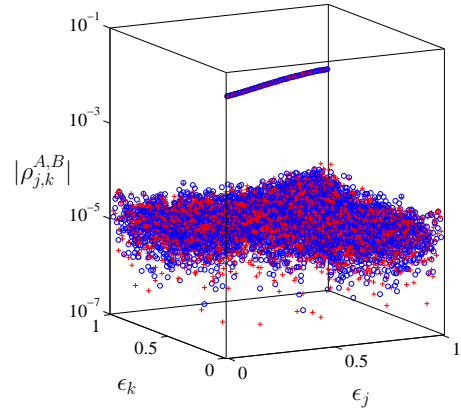}
\caption{(color online). The reduced density matrices of peers with the uniform density of states
after  equilibration is completed. Crosses (red) show the elements of the density matrix of  system $A$ (initially ``hot''),
while open circles (blue) show those for system
$B$ (initially ``cold''). Both peers were initially in canonical thermal states, with the corresponding density matrices
in diagonal form. The parameters are the same as those in Fig.~\ref{fig2}.} \label{fig4}
\end{figure}

To conclude this section, we discuss the important issue of off-diagonal elements of the reduced density
matrices of the peers \textit{after they reached the state of a joint thermal equilibrium}. With
Figs.~\ref{fig1}-\ref{fig3}, we addressed the diagonal elements of the reduced density matrices only and showed that they fit the thermal distributions with the
equilibrium temperature given by Eq.~(\ref{eq:temp_eq}). Remarkably, the off-diagonal elements,  although they appear
during the equilibration process, remain extremely small after  equilibration is completed. Therefore, the thermalized
density matrices of the peers preserve their diagonal forms and remain near canonical, see Fig.~\ref{fig4}.

\section{Thermal equilibration of interacting spin clusters} Synthesized Hamiltonians, although very useful for numerical studies ~\cite{gemmer1},  have a serious  drawback. Namely, they do not feature some nontrivial statistical properties which may be present in spectra of actual quantum systems. Therefore the equipartition scenario needs to be tested with a realistic physical Hamiltonian.

\begin{figure}[t]
\includegraphics[width=0.45\textwidth]{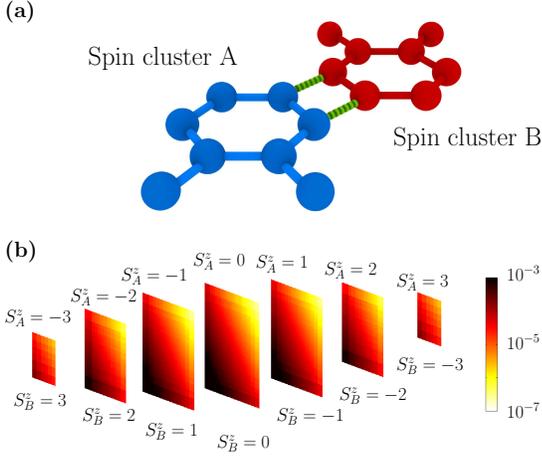}
\caption{(color online). (a) Two locally interacting spin clusters. (b)
The diagonal elements of the density matrices corresponding to the different
magnetization subspaces after the thermalization process is completed.
The Hilbert space of a single cluster splits into $2N_S+1=9$ invariant subspaces,
$\mathcal{H}_{S^z_{X}}$, $X = A, B$, with a  spin $S^z_{X}$ taking  integer values
from $-N_s/2=-4$ to $N_s/2=4$. The interaction between clusters leaves the total
magnetization of the composite system, $S_z$, invariant. Therefore, for the chosen
initial conditions with subspaces $S^z_{A,B}=0$ only populated, the consecutive
equilibration process is restricted to the $S_z=0$ subspace of the Hilbert space of
the composite system. As a result of the interaction, all possible products of local
subspaces, $\mathcal{H}_{S^z_{A}}\otimes\mathcal{H}_{S^z_{B}}$, with opposite magnetization,
$S^z_A = -S^z_B$, become populated. Note that the single-state subspaces with $S_z^{A,B}=\pm4$
are not shown. Initial temperatures of the clusters are the same as in Fig.~\ref{fig2}.
} \label{fig5}
\end{figure}

As a last peer model we use a finite cluster of $N_S=8$ interacting $1/2$-spins. Two clusters are placed into a constant magnetic field, pointing along $z$-direction, and brought into a local contact, see Fig.~\ref{fig5}(a).  Each cluster has $N=2^8=256$ states, so that the overall dimension of the Hilbert space of the composite system is $\mathcal{N}=2^{2N_S}=2^{16}=65536$.

For two identical clusters, we employ here the spin model that is also referred as to XXZ model with the following Hamiltonian, $H_A=H_B\equiv H$:
\begin{equation}
\label{eq:spins}
H = V\sum_{\langle ij \rangle}S_i^zS_j^z-J\sum_{\langle ij \rangle}(S_i^xS_j^x+S_i^yS_j^y)+M\sum_{i} S_i^z,
\end{equation}
where $S^x_i,S^y_i,S^z_i$ are spin-$1/2$ operators on site $i$, $V$ ($J$) are the exchange constants in $z$ ($x$, $y$) directions,
$M$ is the external magnetic field, and $\langle ... \rangle$ indicates here all pairs of next-neighbor spins connected according
to bonds of a single spin cluster displayed in Fig.~\ref{fig5}(a). The coupling term between the clusters, $H^{\rm int}$, assumes
similar to the Hamiltonian of the spin cluster form:
\begin{equation}
\label{eq:spins_int}
H^{\rm int} = V\sum_{\langle ij \rangle_\lambda}S_i^zS_j^z-J\sum_{\langle ij \rangle_\lambda}(S_i^xS_j^x+S_i^yS_j^y).
\end{equation}
Here the sum runs over the two bonds $\langle ij \rangle_\lambda$ that bind the two spin clusters together upon the action of quench.

In distinct contrast to the synthesized model discussed before, both single clusters and the entire composite system  possess integrals of motion additional to the total energy. That are the total magnetization along direction of the applied magnetic field,  $S^z_{A,B}=\sum_{j_{A,B}} S^z_j$, for the clusters, and $z$-component of the total spin, $S^z=\sum_j S^z_j$, for the composite system \cite{sym}.  As a consequence, the Hamiltonian of a single $X$-cluster, $X = A$ or $B$, factorizes over the product space $\bigotimes \mathcal{H}_{S^z_{X}}$ into $2N_S+1$ independent blocks. So does the Hamiltonian of the composite system over the product space $\bigotimes \mathcal{H}_{S^z}$, yielding $4N_S+1$ blocks.

Conservation of the total magnetization allows to study the process of mutual quantum
equilibration  in a more complex situation. For both clusters we choose initial states with only invariant subspaces $S^z_{A,B}=0$ thermally populated. By resorting to the equipartition hypothesis, we predict that a weak interaction quench that preserves the
magnetization of the composite system, $S^z = S^z_A+S^z_B=0$, but violates the separate conservation of the
magnetization of individual cluster, $S^z_{A,B}$, would not only lead to the equilibration between the
subspaces  $S^z_{A,B}=0$, but shall also initiate a population \textit{and consecutive thermalization} within
subspaces  $S^z_{A,B}\neq 0$.

\begin{figure}[t]
\includegraphics[width=0.4\textwidth]{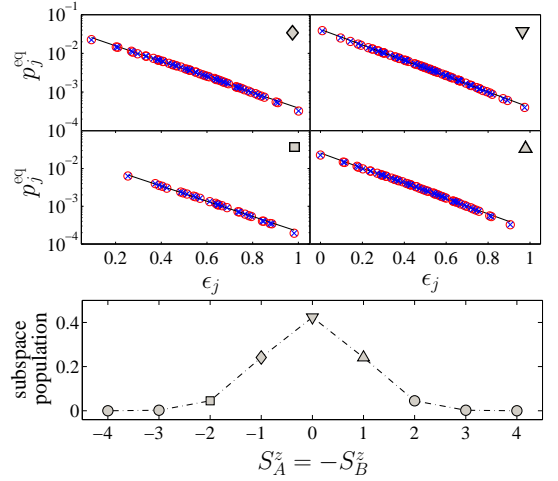}
\caption{(color online). Populations of different $S^z_{A,B}$ subspaces of the Hilbert space of a single cluster
after the equilibration process is completed. Top panels: Equilibrium populations of the energy
levels in subspaces $S^z_A$, being marked by the (blue) crosses, and  $S^z_B = - S^z_A$, as marked by (red) circles, 
are shown for the cluster $A$ and for the cluster $B$, respectively. The solid line for each subspace depicts the thermal energy level populations at
the equilibrium temperature $T_{\rm eq}$, multiplied by the total population of the corresponding subspace. The latter subspace is marked by the filled grey symbol in each panel, as  shown in the bottom part. The equilibrium populations
and the thermal distributions agree (within line thickness) with the analytical prediction. Bottom panel:
Individual population values  of the corresponding subspaces $S^z_A = - S^z_B$ after equilibration. The parameters are the same as in Fig.~\ref{fig5}.}\label{fig6}
\end{figure}

Our analytical calculations based on generalized form of Eqs.~(\ref{eq:step2}-\ref{eq:step3}) for the
factorized space $\mathcal{H}_{S^z=0}=\sum_{S^z_A} \mathcal{H}_{S^z_A}\otimes\mathcal{H}_{-S^z_A}$
perfectly agree with exact diagonalization of the model Hamiltonian in the subspace of zero total magnetization,
$S^z=0$, spanned by $12870$ states, see Figs.~\ref{fig5},~\ref{fig6}. The model parameters are
$J = 0.2 \Delta\epsilon$, $V = 0.1 \Delta\epsilon$, $M = 0.05 \Delta\epsilon$, $\lambda_{\rm int}
= 0.095(\bar s/\bar h)=1$.

The equilibrium temperature $T_{\rm eq}$ was calculated by using Eq.~(\ref{eq:temp_eq}),
which was applied to the initially populated subspace, $\mathcal{H}_{S^z_{A}=0}\otimes\mathcal{H}_{S^z_{B}=0}$, only.
It is noteworthy that the 'equilibrium' distributions for different subspaces  perfectly match the thermal distributions
with the same equilibrium temperature, $T_{\rm eq}$, see in Fig.~\ref{fig6} (top panels).

\section{Summary and outlook} In conclusion, using different classes of Hamiltonians, we
have unraveled the mechanism responsible for the thermal equilibration of two identical quantum
peers prepared initially in canonical states at different temperatures. This mechanism, i.e., the
equipartition within energy shells in the Hilbert space of the composite system, may appear
whenever the interaction is small enough to satisfy the weak-coupling condition, given by 
Eqs.~(\ref{eq:weak coupling}, \ref{eq:not-to-weak coupling}).
However, the equipartition scenario is not universal: Quantum systems that exhibit Anderson
localization are expected to invalidate the equipartition scenario when coupled by a weak local interaction,
and the final equilibrium states of the corresponding peers can differ substantially from being thermal-like~\cite{huse,gogolin}.

One should keep in mind that the time evolution of any isolated quantum system with a finite number of levels has a finite recurrence time,
which depends on the system spectrum and the system initial conditions. Thus the equilibration of the peers to a thermal `equilibrium'
after some interaction time $t$ does not contradict the disappearance of the equilibration at some larger times, $t_{\rm rec} > t$, due to
revivals. The revival time scales can be very short when the interacting systems are small~\cite{larson}.

The equilibration process is governed by the Hamiltonians, $H_A$, $H_B$, and $H^{\rm int}$, and its output is in one-to-one correspondence with the initial states of the peers. It means that initial states different from thermal Gibbs states, generally would lead to a final quasi-equilibrium which may not be thermal-like anymore. This complication, however, could be weakened by the increase of the number of peers: interaction between $M \gg 2$ systems  would effectively mimic an environment for a single peer, thus leading to the mutual equilibration of all peers to nearly identical thermal states regardless the shape of their initial eigenstate distributions~\cite{bolz, therm_inf_d}.

We acknowledge the support by the German Excellence Initiative ``Nanosystems Initiative Munich
(NIM)''.


\begin{thebibliography}{1000}
\bibitem{quasi} In a  mathematical sense, there is no unidirectional relaxation towards a strict stationary  state
in a closed quantum system. For any finite quantum system with a discrete energy spectrum, the evolution of any observable is quasiperiodic, and, therefore, recurrences are inevitable, see  in Ref. [22].
However, when the dimension of the system Hilbert space is sufficiently large, the recurrences
occur on time scales which are much longer than any time scale of practical relevance [11].
\bibitem{esposito} The microcanonical equipartition inside the energy shell,
initiated by coupling of a system to an environment of finite heat capacity,
has been considered in~\cite{esposito1}.
\bibitem{sym} In addition, each cluster has also a mirror symmetry
so that the corresponding Hilbert space is shared by symmetric and antisymmetric
eigenstates. However, in the context of mutual thermalization
this parity can be safely left without paying further attention to it.
\end{thebibliography}
\end{document}